\documentclass{article}
\usepackage{spconf,amsmath,graphicx,hyperref,amssymb,booktabs,float,tabularx,xcolor,soul}
\usepackage{algorithm}
\usepackage{algpseudocode}

\sethlcolor{lime}

\title{Noise-Robust Contrastive Learning with an MFCC-Conformer For Coronary Artery Disease Detection}

\name{Milan Marocchi$^*$, Matthew Fynn$^*$, Yue Rong}

\address{Curtin University, Bentley
6102, WA, Australia}

\begin{document}

\maketitle
\renewcommand\thefootnote{}\footnotetext{$^*$The first two authors contributed equally.} 
\footnotetext{\noindent This paper has been accepted for presentation at ICASSP 2026. 
© 2026 IEEE. Personal use of this material is permitted. Permission from IEEE must be obtained for all other uses, in any current or future media, including reprinting/republishing this material for advertising or promotional purposes, creating new collective works, for resale or redistribution to servers or lists, or reuse of any copyrighted component of this work in other works.}
\begin{abstract}
Cardiovascular diseases (CVD) are the leading cause of death worldwide, with coronary artery disease (CAD) comprising the largest subcategory of CVDs. Recently, there has been increased focus on detecting CAD using phonocardiogram (PCG) signals, with high success in clinical environments with low noise and optimal sensor placement. Multichannel techniques have been found to be more robust to noise; however, achieving robust performance on real-world data remains a challenge. This work utilises a novel multichannel energy-based noisy-segment rejection algorithm, using heart and noise-reference microphones, to discard audio segments with large amounts of non-stationary noise before training a deep learning classifier. This conformer-based classifier takes mel-frequency cepstral coefficients (MFCCs) from multiple channels, further helping improve the model's noise robustness. The proposed method achieved 78.4\% accuracy and 78.2\% balanced accuracy on 297 subjects, representing improvements of 4.1\% and 4.3\%, respectively, compared to training without noise-segment rejection.
\end{abstract}
\vspace*{-4mm}
\section{Introduction}
\label{sec:intro}

Cardiovascular disease (CVD) result in 31\% of deaths annually around the globe~\cite{who}. Coronary artery disease (CAD) is the largest subtype. CAD requires prompt diagnosis to help manage the disease before it progresses. However, auscultation
yields relatively low diagnostic accuracy, partly because heart sounds often lie near the threshold of
human hearing~\cite{heart1, heart2, heart3}. With expensive and highly invasive angiograms being the gold-standard diagnostic tool~\cite{cadangiogram}, there is a gap in prescreening tools. Recently deep learning aided phonocardiogram (PCG) methods have been employed to accuractely pre-screen CAD~\cite{vestpaper, arnab}. It has been found that multichannel PCG signals help improve CAD classification performance in the presence of background hospital noise~\cite{pathak}. The use of linear frequency cepstral coefficients (LFCCs) and mel-frequency cepstral coefficients (MFCCs) have further been found to improve the noise robustness of deep learning models~\cite{vestpaper}. Modern strong performing transformer-based architectures such as conformers have seen success in speech but have not yet been evaluated in PCG signals~\cite{conformer}.
They have been found to lead to state-of-the-art (SOTA) performance for speech in clean and noisy conditions, showing promise for use on noisy PCG data.
However, there is currently a lack of work utilising all these techniques, along with noisy segment removal, based on external and internal noise, on a real-world noisy dataset to validate all of these approaches~\cite{pcg_survey}. This study makes use of data collected in a noisy hospital environment with unoptimal sensor placement, to evaluate the performance of noise-segment rejection, along with a conformer-based model that uses MFCCs as input. 

This work's novel contributions to the field are:
\begin{itemize}
    \vspace*{-2mm}
    \item A multichannel noise-aware signal conditioning method that jointly rejects noisy PCG segments using energy cues from both the heart and noise-reference microphones
    \vspace*{-2mm}
    \item An integrated MFCC–Conformer pipeline with supervised hybrid-contrastive learning that, together with the conditioning stage, results in robust and balanced CAD detection performance in real-world noisy settings
\end{itemize}
\vspace*{-5mm}
\section{MATERIALS}
\label{sec:format}
All data processing and model training were conducted using a Ryzen 7 3800X CPU and an Nvidia RTX 3090 (24~GB), with Python 3.11 and PyTorch 2.1.2.

\subsection{Data Aquistion}
A wearable vest embedded with multiple PCG sensors was used to acquire synchronised multichannel PCG data from participating subjects \cite{vestpaper}. Each stethoscope incorporated two microphones: one positioned beneath the diaphragm (Heart mic – HM) and another on the rear of the stethoscope (Noise mic – NM). The vest can be fitted easily, requiring less than a minute. This work made use of channels 1, 2, 3 and 4 of the seven PCG channels.
\vspace{-2mm}
\subsection{Dataset}
The wearable vest collected data from 297 male subjects at Fortis Hospital, Kolkata, across three separate rounds: May–June 2023, January–February 2024, and February 2025. Of these, 155 subjects were diagnosed with CAD, defined as having greater than 50\% stenosis in the right coronary artery, left coronary artery, or left circumflex artery, confirmed through angiography. The remaining 142 subjects were classified as normal, including 32 subjects under 35 years of age, assumed to be free of CAD as the risk is significantly higher in males above 45 years \cite{hajar2017risk}. Data was collected in a clinical environment, and thus typical hospital background noise was present, including talking, privacy curtains closing, and doors slamming. Subjects were instructed to sit comfortably on a chair and breathe normally during recording. Between one and three 60-second recordings were acquired from each subject. 



\vspace*{-2mm}
\section{METHOD}
\label{sec:pagestyle}

Segments of audio from the PCG signals are extracted and preprocessed before being used to train a conformer-based classifier with a contrastive loss. The methods will first detail the novel energy-based noisy segment rejection approach, preprocessing, and feature extraction before detailing the model training and inference.
\vspace{-3mm}
\subsection{Preprocessing}
The PCG signals first are concatenated so that there is one contiguous recording for each subject. This will ensure that there is no data leakage. The regions around the joins are then discarded when segmenting the signal, which is further discussed in Section~\ref{sec:segment}. Following this, the signals undergo noisy segment rejection, which will mark noisy segments so that they will not be included in any fragments. Following this, the signals undergo spike removal \cite{schmidt} and then are bandpassed using a second-order Butterworth filter between 25Hz and 450Hz. The signals are then k-peak mean normalised \cite{arnab}, before being used to extract features. Then the signals are segmented into fragments to be used for training.

\vspace*{-3mm}
\subsection{Noisy Segment Rejection}
\label{sec:denoise}
\begin{figure}
    \centering
    \includegraphics[width=0.85\linewidth]{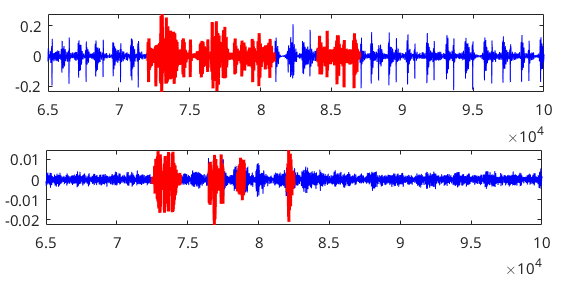}
    \caption{\footnotesize HM (top) and NM (bottom) signals with noise-corrupted segments highlighted in red, which were discarded from all channels downstream.}
    \label{imp fig}
    \vspace*{-6mm}
\end{figure}

Both HM and NM signals were utilised for noisy segment rejection. We propose an algorithm that identifies and mitigates impulsive and movement noise within the recordings. The algorithm outputs the set of indices deemed corrupted by impulse noise, enabling clean signal segments to be used in downstream training and inference. The algorithm takes a signal (either from HM or NM) as input and divides it into frames of fixed length. For each frame, the energy is computed as the sum of squared samples. The median frame energy (excluding the first and last frames) is then calculated. Each frame whose energy exceeds the product of the median value and the given threshold is flagged, and the corresponding start and end indices of that frame are stored in a variable. Additionally, the first and last seconds of each recording are flagged as noisy, ensuring boundaries between concatenated signals are excluded from downstream tasks, including filtering. Algorithm \ref{alg:medpow} describes this process of highlighting noisy segment indices. Sources of identifiable noise included sudden bursts of external voices and door slams, while patient movement introduced friction noise between the diaphragm and the skin. As the NM signals from all stethoscopes were highly correlated, only channel 4 was used to detect noisy indices, whereas each HM was processed separately. The frame length was tailored to the dominant noise source of each channel: for HM signals, it was set to 2.5 s to capture longer-duration friction noise, while for NM signals, it was set to 0.25 s to detect brief impulsive events such as door slams or speech bursts. For both signal types, the threshold was fixed at 2.5 times the median frame energy, chosen to balance sensitivity to noise events against robustness to natural signal variability. Figure~\ref{imp fig} illustrates an example HM and NM signal with noisy indices highlighted in red. The resulting indices from all HM and NM signals were then combined to form a final vector of noise-corrupted segments across the concatenated recording, accounting for overlapping indices highlighted by separate channels. From this combined output, the complementary indices corresponding to noise-free segments were extracted and applied uniformly across the entire multichannel recording. 

\begin{algorithm}[t]
\scriptsize
\caption{Noisy Segment Identification}
\label{alg:medpow}
\begin{algorithmic}[1]
\Require Signal $x[0\ldots L-1]$, sampling rate $f_s$, frame length (s) $T_f$, threshold $\tau$
\Ensure List of index intervals $\mathcal{I}$ containing frames flagged as noise
\State $N \gets \lfloor L / (T_f \cdot f_s) \rfloor$ \Comment{number of full frames}
\State $F \gets T_f \cdot f_s$ \Comment{samples per frame}
\State $\mathbf{E} \gets \mathbf{0}_{1 \times N}$ \Comment{frame energies}
\For{$i = 0$ to $N-1$}
    \State $s \gets iF$; \quad $e \gets (i+1)F -1$
    \State $\mathbf{E}[i] \gets \sum_{n=s}^{e} x[n]^2$ \Comment{sum of squares (energy)}
\EndFor
\State $m \gets \mathrm{median}\big(\mathbf{E}[1{:}N\!-\!2]\big)$ \Comment{exclude first/last frame}
\State $\mathcal{I} \gets [\ ]$ \Comment{empty list of (start,end) indices}
\For{$i = 1$ to $N$}
    \If{$\mathbf{E}[i] > \tau \cdot m$}
        \State $s \gets iF $; \quad $e \gets (i+1)F-1$
        \State append $(s,e)$ to $\mathcal{I}$
    \EndIf
\EndFor
\State \Return $\mathcal{I}$
\end{algorithmic}
\end{algorithm}
\vspace*{-5mm}

\subsection{Segmentation}
\label{sec:segment}

Following denoising, the noise-free indices of each subject’s recording were used to extract clean segments, whose lengths varied according to the distribution of noise. Segments shorter than four seconds were discarded. From the remaining segments, fixed-length fragments of four seconds were extracted. To ensure class balance during training, a base number of fragments $F_\text{base}$ was specified, with additional fragments drawn from the underrepresented class until equal counts were obtained. In this case, with more CAD subjects ($N_\text{CAD}$) than normal subjects ($N_\text{NOR}$), the class-level fragment counts were set as
\begin{equation}
F_\text{CAD} = F_\text{base},
\qquad
F_\text{NOR} = \frac{N_\text{CAD}}{N_\text{NOR}} \cdot F_\text{base}
\end{equation}
During validation and testing, the target number of fragments was fixed across classes to avoid bias. For each subject, the assigned number of fragments $F_\text{class}$ ($F_\text{CAD}$ or $F_\text{NOR}$) was distributed across its noise-free segments in proportion to segment duration. Specifically, for the $i^\text{th}$ noise-free segment,
\begin{equation}
F_i = \left\lfloor F_\text{class} \cdot \frac{L_i}{\sum_j L_j} \right\rfloor,
\end{equation}
where $L_i$ is the length of the $i^\text{th}$ segment and $\sum_j L_j$ is the total length of all valid segments for that subject. Any remaining fragments were allocated to the longest segments to preserve proportionality, ensuring $\sum_i F_i = F_\text{class}$. Finally, fragments were extracted with variable overlap, determined by both $F_i$ and the length of the corresponding segment.
\vspace{-3mm}
\subsection{Feature Extraction}
MFCCs were extracted from each recording segment following amplitude normalisation to mitigate inter-recording variability. The normalisation conducted was k-peak normalisation, which has been shown to be more effective for PCG signals~\cite{arnab}. For each segment, the short-time Fourier transform (STFT) was computed, mapped to the mel scale, and subsequently transformed into cepstral coefficients. There were 128 MFCCs extracted between 25Hz and 450Hz with a window length of 512 samples, hop length of 160 samples for the STFT computation. The MFCC vectors obtained from individual segments for each channel were then concatenated along the time channel axis to create a unified representation of all features from each channel.

\vspace{-3mm}
\subsection{Models}
The proposed network is a conformer-style encoder operating on MFCC sequences. After per-segment MFCC preprocessing (single- or 
multi-channel; features concatenated across channels), a linear projection maps the 
input $F$-dimensional frames to the model width $D$. The encoder comprises $B$ 
stacked conformer blocks, each following a layered topology: a pre-normalized 
feed-forward sublayer (scaled by $0.5$), multi-head self-attention with $H$ heads, a convolutional 
module, and a second pre-normalised feed-forward sublayer of dimension $M$ (again scaled by $0.5$), 
with residual connections throughout. The convolutional module employs a pointwise expansion with gated linear units (GLU), 
depthwise convolution (kernel size $k$, which was fixed to 31), batch normalisation, Sigmoid linear unit (SiLU) activation, and a 
pointwise projection back to $D$. Layer normalisation is applied before attention and 
convolutional sublayers, and dropout is used in the feed-forward paths. A final layer 
normalisation precedes temporal aggregation via adaptive average pooling to produce 
an embedding, which is fed to a shallow MLP classifier (one hidden 
layer with ReLU and dropout) to predict one of the two classes. 


\begin{table}[!h]
\scriptsize
\centering
\begin{tabular}{ll|ll}
\toprule
\textbf{Parameter} & \textbf{Value} & \textbf{Parameter} & \textbf{Value} \\
\midrule
batch size ($N_{b}$)   & 256       & $\alpha$      & 0.7235 \\
learning rate & 2.97e-06  & $\beta$       & 0.9807 \\
weight decay & 5.71e-05  & temperature   & 0.8050 \\
$s$          & 2         & $\lambda_c$   & 0.00281 \\
$\gamma$     & 0.2903    & $D$           & 1024 \\
$H$          & 8         & $M$           & 128 \\
$B$          & 3         & dropout & 0.2903 \\
\bottomrule
\end{tabular}
\caption{Hyperparameter values.}
\label{tab:hpo}
\end{table}

\begin{table*}[ht]
\scriptsize
\centering
\caption{Model performance at the fragment and subject level - Base number of fragments is 61}
\label{tab:frag_subj}
\begin{tabularx}{\textwidth}{p{3cm} l l l l l l l}
\toprule
\textbf{Method } & \textbf{Acc} & \textbf{UAR} & \textbf{TPR} & \textbf{TNR} & \textbf{F1$^+$} & \textbf{F1$^-$} & \textbf{MCC} \\
\midrule
\multicolumn{8}{c}{\textbf{Fragment Level}} \\
\midrule
Noisy MFCC Conformer & 71.2$\pm$0.05\% & 70.9$\pm$0.05\% & 77.5$\pm$0.07\% & 64.2$\pm$0.17\% & 73.9$\pm$0.02\% & 67.5$\pm$0.09\% & 0.425$\pm$0.009 \\
Denoised MFCC Conformer & \textbf{73.9$\pm$0.35\%} & \textbf{73.7$\pm$0.37\%} & 76.8$\pm$0.42\% & \textbf{70.7$\pm$0.76\%} & 75.5$\pm$0.32\% & \textbf{71.9$\pm$0.44\%} & \textbf{0.478$\pm$0.072} \\
Noisy Wav2Vec 2.0~\cite{milanwav2vec} & 70.7$\pm$0.21\% & 68.2$\pm$0.18\% & \textbf{78.9$\pm$0.46\%} & 57.6$\pm$0.30\% & \textbf{76.8$\pm$0.22\%} & 59.9$\pm$0.22\% & 0.372$\pm$0.004 \\
\midrule
\multicolumn{8}{c}{\textbf{Subject Level}} \\
\midrule
Noisy MFCC Conformer & 74.3$\pm$0.09\% & 73.9$\pm$0.10\% & 80.9$\pm$0.11\% & 66.9$\pm$0.30\% & 76.8$\pm$0.06\% & 70.6$\pm$0.15\% & 0.490$\pm$0.019 \\
Denoised MFCC Conformer & \textbf{78.4$\pm$0.29\%} & \textbf{78.2$\pm$0.32\%} & 81.9$\pm$0.49\% & \textbf{74.5$\pm$0.97\%} & 79.9$\pm$0.20\% & \textbf{76.4$\pm$0.48\%} & \textbf{0.570$\pm$0.058} \\
Noisy Wav2Vec 2.0~\cite{milanwav2vec} & 77.1$\pm$1.50\% & 74.3$\pm$1.73\% & \textbf{86.5$\pm$1.30\%} & 62.0$\pm$2.76\% & \textbf{82.3$\pm$1.10\%} & 67.1$\pm$2.56\% & 0.510$\pm$0.035 \\
\bottomrule
\end{tabularx}
\end{table*}

\vspace*{-7mm}
\subsection{Contrastive Learning}
\vspace{-2mm}
\subsubsection{Hybrid-Contrastive Loss} 
A hybrid-contrastive loss is utilised to best shape the embedding space to ensure more robust classifications, especially in the case of noise, which is typical within this dataset. It is a supervised loss function to best make use of the data being fully labelled. The training objective combines three components: a supervised contrastive loss, 
a classification loss, and an optional center loss. Given a batch of feature 
embeddings $\mathbf{z}_i \in \mathbb{R}^d$ with corresponding class labels $y_i$, 
the total loss is defined as
\begin{equation}
\mathcal{L} = 
\beta \, \mathcal{L}_{\text{contr}}(\mathbf{z}, \mathbf{y}; \tau) \;+\; 
\alpha \, \mathcal{L}_{\text{CE}}(\mathbf{p}, \mathbf{y}) \;+\; 
\lambda_c \, \mathcal{L}_{\text{center}}(\mathbf{z}, \mathbf{y}),
\end{equation}
where $\mathbf{p}$ are the classifier logits, $\alpha$ and $\beta$ weight the 
classification and contrastive terms, respectively, and $\lambda_c$ controls the 
influence of the center loss. The hyperparameter $\tau$ denotes the temperature. A standard cross-entropy objective is used for the classification loss ($\mathcal{L}_{\text{CE}}$).
\vspace*{-4mm}
\subsubsection{Supervised Contrastive Loss} 
We normalise all embeddings and compute a cosine similarity matrix 
$\mathbf{S} = \mathbf{Z} \mathbf{Z}^{\top}$. 
Where $\mathbf{Z}$ is the batch of normalised feature embeddings; $\mathbf{Z} = [\mathbf{z}_1, ..., \mathbf{z}_n]$. The contrastive loss 
encourages embeddings of the same class to be close, and embeddings of different 
classes to be pushed apart:
\begin{equation}
\mathcal{L}_{\text{contr}} = -\frac{1}{N_{mb}} \sum_{i=1}^{N_{mb}} 
\frac{1}{|\mathcal{P}(i)|} \sum_{j \in \mathcal{P}(i)} 
\log \frac{\exp(\mathbf{S}_{ij}/\tau)}
{\sum_{k=1}^{N_{mb}} \exp(\mathbf{S}_{ik}/\tau)},
\end{equation}
where $\mathcal{P}(i)$ is the set of positive indices (samples with $y_j = y_i$, 
excluding $i$ itself), and $N_{mb}$ is the number of samples within the mini batch.
\vspace*{-4mm}
\subsubsection{Center Loss} 
For each class $c$, a learnable center vector $\mathbf{c}_c \in \mathbb{R}^d$ is 
maintained. The center loss penalises the distance between feature vectors and 
their class centers:
\begin{equation}
\mathcal{L}_{\text{center}} = 
\frac{1}{N_{mb}} \sum_{i=1}^{N_{mb}} \lVert \mathbf{z}_i - \mathbf{c}_{y_i} \rVert_2^2.
\end{equation}

\vspace*{-0.5cm}
\subsection{Model Training}
Training is done on a fragment level to optimise the fragment-level metrics. The model was trained using the AdamW optimiser~\cite{adamw}. An exponential decay learning rate scheduler was also utilised parameterised by the step size ($s$) and the decay rate ($\gamma$). The model makes use of gradient accumulation with a mini batch size ($N_{mb}$) and a batch size ($N_{b}$), where the gradients from each mini batch update are accumulated until the number of samples sum to the $N_{b}$. The models are trained for 10 epochs, with the best model from training being selected as the one with the best weighted average Matthew's Correlation Coefficient (MCC) between the training and validation set; a scaling factor of 0.9 to the validation MCC and 0.1 to the training MCC. The MCC metric provides a single measure that captures all aspects of model performance~\cite{mcc}. For tuning the hyperparameters and the architecture of the model, a Bayesian optimisation was conducted using the Optuna library~\cite{optuna}. Table~\ref{tab:hpo} contains the parameters included in this optimisation. Each trial was repeated three times and was optimised over the average validation MCC score to ensure a less noisy value being used.  









\vspace*{-0.35cm}
\subsection{Model Inference}
Before being used for inference, the MLP is removed and replaced with a support vector machine (SVM) with a radial basis function (RBF) kernel. The subject-level predictions are then taken by majority vote of each of the fragment-level predictions. The accuracy, unweighted average recall (UAR), true positive rate (TPR), true negative rate (TNR), F1 scores, and MCC are reported.
\vspace*{-2mm}
\section{RESULTS AND DISCUSSION}
\label{sec:typestyle}

Table~\ref{tab:frag_subj} displays the fragment and subject performance which compares the baseline with no noise-segment rejection to a model that was trained with the contrastive loss and the signals denoised. These results are presented as average$\pm$standard deviation, where the models are averaged over the five folds and run three times to account for the stochasticity of the training of the neural networks. The table also contained a comparison to a previous method on this same vest data, with the best method for each metric being highlighted in bold.

The proposed noise-segment rejection algorithm, when applied in conjunction with the conformer-based model, yielded subject-level improvements of 4.1\%, 4.3\%, and 0.08 in accuracy, UAR, and MCC, respectively. The model's performance was also significantly more balanced between TPR and TNR with the use of the noise-segment rejection, highlighting the importance of removing segments heavily contaminated with non-stationary noise. Comparing this method to another work utilising data from the same vest, it is seen that this method results in a more performant and noise robust model, with subject-level increases of 1.3\%, 3.9\% and 0.06 in the accuracy, UAR and MCC, respectively. This confirms that the use of these techniques to deal with a noisy real-world dataset help to improve performance, whilst also providing a model which is signifigicantly smaller than the one in \cite{milanwav2vec}. It is much smaller as an early feature fusion is employed, as opposed to a late feature fusion.

\vspace*{-1mm}
\section{CONCLUSION AND FURTHER WORK}
This work detailed an end-to-end CAD classification pipeline that integrates noise-aware segment rejection with multichannel MFCC–Conformer modelling and hybrid contrastive learning, yielding more robust and balanced performance on noisy PCG data than a previous Wav2Vec-based method. Future work will include ablations and cross-dataset experiments to better quantify component contributions and generalisation.

\label{sec:majhead}
\textit{\textbf{Acknowledgment:}} — We would like to thank Ticking Heart Pty Ltd for providing the wearable vest design, and Fortis Hospital Kolkata and Dr. Mandana for their support in data collection. We would also like to thank Danny Baker, Tank and Enzo for their support during the development of this work.

\textit{\textbf{Code Availability:}} — All code is available at {\url{https://github.com/MilanMarocchi/noise-robust-cad-conformer}}

\textit{\textbf{Ethics approval:}} — This study received approval from the ethics committee of Fortis Hospital, Kolkata, India, where the data collection took place (ECR/240/Inst/WB/2013/RR-19, Date of approval: 13/01/2023) in accoradance to the Helsinki Declaration. Informed consent was obtained from all subjects.



\bibliographystyle{IEEEtran}
\bibliography{IEEEabrv, refs}

\begin{thebibliography}{10}
\providecommand{\url}[1]{#1}
\csname url@samestyle\endcsname
\providecommand{\newblock}{\relax}
\providecommand{\bibinfo}[2]{#2}
\providecommand{\BIBentrySTDinterwordspacing}{\spaceskip=0pt\relax}
\providecommand{\BIBentryALTinterwordstretchfactor}{4}
\providecommand{\BIBentryALTinterwordspacing}{\spaceskip=\fontdimen2\font plus
\BIBentryALTinterwordstretchfactor\fontdimen3\font minus \fontdimen4\font\relax}
\providecommand{\BIBforeignlanguage}[2]{{%
\expandafter\ifx\csname l@#1\endcsname\relax
\typeout{** WARNING: IEEEtran.bst: No hyphenation pattern has been}%
\typeout{** loaded for the language `#1'. Using the pattern for}%
\typeout{** the default language instead.}%
\else
\language=\csname l@#1\endcsname
\fi
#2}}
\providecommand{\BIBdecl}{\relax}
\BIBdecl

\bibitem{who}
WHO, "Cardiovascular Diseases (CVDs)". \emph{Geneva, Switzerland: WHO}, 2021.

\bibitem{heart1}
M.~A. Chizner, ``Cardiac auscultation: {{Rediscovering}} the lost art,'' \emph{Current Problems in Cardiology}, vol.~33, no.~7, pp. 326--408, Jul. 2008.

\bibitem{heart2}
C.~A. Feddock, ``The {Lost Art} of clinical skills,'' \emph{The American Journal of Medicine}, vol. 120, no.~4, pp. 374--378, Apr. 2007.

\bibitem{heart3}
Q.-M. Zhao, C.~Niu, F.~Liu, L.~Wu, X.-J. Ma, and G.-Y. Huang, ``Accuracy of cardiac auscultation in detection of neonatal congenital heart disease by general paediatricians,'' \emph{Cardiology in the Young}, vol.~29, no.~5, pp. 679--683, May 2019.

\bibitem{cadangiogram}
\BIBentryALTinterwordspacing
R.~J. Gibbons, K.~Chatterjee, J.~Daley, J.~S. Douglas, S.~D. Fihn, J.~M. Gardin, M.~A. Grunwald, D.~Levy, B.~W. Lytle, R.~A. O’Rourke, W.~P. Schafer, S.~V. Williams, J.~L. Ritchie, R.~J. Gibbons, M.~D. Cheitlin, K.~A. Eagle, T.~J. Gardner, A.~Garson, R.~O. Russell, T.~J. Ryan, and S.~C. Smith, ``Acc/aha/acp-asim guidelines for the management of patients with chronic stable angina1: A report of the american college of cardiology/american heart association task force on practice guidelines (committee on management of patients with chronic stable angina),'' \emph{Journal of the American College of Cardiology}, vol.~33, no.~7, pp. 2092--2197, 1999. [Online]. Available: \url{https://www.sciencedirect.com/science/article/pii/S0735109799001503}
\BIBentrySTDinterwordspacing

\bibitem{vestpaper}
M.~Fynn, K.~Mandana, J.~Rashid, S.~Nordholm, Y.~Rong, and G.~Saha, ``Practicality meets precision: Wearable vest with integrated multi-channel pcg sensors for effective coronary artery disease pre-screening,'' \emph{Computers in Biology and Medicine}, vol. 189, p. 109904, 2025.

\bibitem{arnab}
\BIBentryALTinterwordspacing
A.~Maity and G.~Saha, ``Enhancing cross-domain robustness in phonocardiogram signal classification using domain-invariant preprocessing and transfer learning,'' \emph{Computer Methods and Programs in Biomedicine}, vol. 257, p. 108462, 2024. [Online]. Available: \url{https://www.sciencedirect.com/science/article/pii/S0169260724004553}
\BIBentrySTDinterwordspacing

\bibitem{pathak}
\BIBentryALTinterwordspacing
A.~Pathak, P.~Samanta, K.~Mandana, and G.~Saha, ``An improved method to detect coronary artery disease using phonocardiogram signals in noisy environment,'' \emph{Applied Acoustics}, vol. 164, p. 107242, 2020. [Online]. Available: \url{https://www.sciencedirect.com/science/article/pii/S0003682X19305742}
\BIBentrySTDinterwordspacing

\bibitem{conformer}
\BIBentryALTinterwordspacing
A.~Gulati, J.~Qin, C.-C. Chiu, N.~Parmar, Y.~Zhang, J.~Yu, W.~Han, S.~Wang, Z.~Zhang, Y.~Wu, and R.~Pang, ``Conformer: Convolution-augmented transformer for speech recognition,'' 2020. [Online]. Available: \url{https://arxiv.org/abs/2005.08100}
\BIBentrySTDinterwordspacing

\bibitem{pcg_survey}
Z.~Ren, Y.~Chang, T.~T. Nguyen, Y.~Tan, K.~Qian, and B.~W. Schuller, ``A comprehensive survey on heart sound analysis in the deep learning era,'' 2023.

\bibitem{hajar2017risk}
R.~Hajar, ``Risk factors for coronary artery disease: historical perspectives,'' \emph{Heart views}, vol.~18, no.~3, pp. 109--114, 2017.

\bibitem{schmidt}
S.~E. Schmidt, C.~{Holst-Hansen}, J.~Hansen, E.~Toft, and J.~J. Struijk, ``Acoustic features for the identification of coronary artery disease,'' \emph{{IEEE} Transactions on Biomedical Engineering}, vol.~62, no.~11, pp. 2611--2619, Nov. 2015.

\bibitem{milanwav2vec}
\BIBentryALTinterwordspacing
M.~Marocchi, M.~Fynn, K.~Mandana, and Y.~Rong, ``Scaling to multimodal and multichannel heart sound classification: Fine-tuning wav2vec 2.0 with synthetic and augmented biosignals,'' 2025. [Online]. Available: \url{https://arxiv.org/abs/2509.11606}
\BIBentrySTDinterwordspacing

\bibitem{adamw}
\BIBentryALTinterwordspacing
I.~Loshchilov and F.~Hutter, ``Decoupled weight decay regularization,'' 2019. [Online]. Available: \url{https://arxiv.org/abs/1711.05101}
\BIBentrySTDinterwordspacing

\bibitem{mcc}
D.~Chicco and G.~Jurman, ``The advantages of the matthews correlation coefficient ({MCC}) over f1 score and accuracy in binary classification evaluation,'' \emph{BMC Genomics}, vol.~21, no.~6, 2020.

\bibitem{optuna}
T.~Akiba, S.~Sano, T.~Yanase, T.~Ohta, and M.~Koyama, ``Optuna: A next-generation hyperparameter optimization framework,'' in \emph{Proceedings of the 25th {ACM} {SIGKDD} International Conference on Knowledge Discovery and Data Mining}, 2019.

\end{thebibliography}

\end{document}